\documentclass[12pt]{article}

\usepackage{epsfig,amsmath,amscd,amssymb,graphicx}

\makeatletter
\@addtoreset{equation}{section}
\renewcommand{\theequation}{\thesection.\arabic{equation}}
\renewcommand{\thefootnote}{\fnsymbol{footnote}}
\makeatother

\newcommand {\beq}{\begin{eqnarray}}
\newcommand {\eeq}{\end{eqnarray}}
\newcommand {\non}{\nonumber\\}

\newcommand {\1}[1]{\frac{1}{#1}}

\newcommand {\thb}{\bar{\theta}}

\newcommand {\ph}{\varphi}

\newcommand {\sig}{\sigma}

\newcommand {\Sig}{\Sigma}

\newcommand {\del}{\partial}
\newcommand {\dagg}{^{\dagger}}

\newcommand {\tr}{{\rm tr}\,}

\newcommand{\hs}[1]{\hspace{#1 mm}}

\begin{document}

%%%%%%%%%%%%%%%%%%%%%%%%%%%%%%%%%
\thispagestyle{empty}
\begin{flushright}
% TIT/HEP-504 \\
{\tt hep-th/0401084} \\
January, 2004 \\
\end{flushright}
\vspace{2mm}
\begin{center}
{\Large
{\bf BPS Domain Walls in Massive Hyper-K\"ahler Sigma Models
} 
%\\
%\vspace{2mm}{\bf 
%}
} 
\\[8mm]
\vspace{3mm}

\normalsize
  {\large \bf 
  Masato~Arai~$^{a}$}
\footnote{\it  e-mail address: 
arai@fzu.cz}
,  
 {\large \bf 
Muneto~Nitta~$^{b}$}
\footnote{\it  e-mail address: 
nitta@th.phys.titech.ac.jp

Address after December 11: Tokyo Institute of Technology
},
~and~~  {\large \bf 
Norisuke~Sakai~$^{c}$}
\footnote{\it  e-mail address: 
nsakai@th.phys.titech.ac.jp
}

\vskip 1.5em

{ \it $^{a}$Institute of Physics, AS CR, 
  182 21, Praha 8, Czech Republic \\
  $^{b}$Department of Physics, Purdue University, 
West Lafayette, IN 47907-1396, USA\\
and \\
  $^{c}$
Department of Physics, Tokyo Institute of 
Technology \\
Tokyo 152-8551, JAPAN   }
\vspace{3mm}
{\bf Abstract}\\[5mm]
{\parbox{13cm}{\hspace{5mm}
%%%%%%%%%%%%%%%%%%%%%%%%%%%%%%%%%%%%%%%%%%%%%%%%%%
%%%%%%%%%%

We discuss hypermultiplets admitting degenerate discrete vacua 
and BPS domain walls interpolating them.
This talk is based on the original papers,
hep-th/0307274, hep-th/0211103 and hep-th/0302028.

%%%%%%%%%%%%%%%%%%%%%%%%%%%%%%%%%%%%%%%%%%%%%%%%%%
%%%%%%%%%%
}}
\end{center}
\vfill
\newpage
\setcounter{page}{1}
\setcounter{footnote}{0}
\renewcommand{\thefootnote}{\arabic{footnote}}

%%%%%%%%%%%%%%%%%%%%%%%%%%%%%%%%%%%%%%%%%%%%%%%%%%%%%%%%%%%%%
% The main text of your paper                               %
%%%%%%%%%%%%%%%%%%%%%%%%%%%%%%%%%%%%%%%%%%%%%%%%%%%%%%%%%%%%%

\section{Introduction}
There exists a long history for searching 
topological defects such as domain walls. 
Lots of their properties in particular localization 
of massless or light fields on them have been investigated. 
Recent interest in the brane-world scenario~\cite{brane} 
has brought us an idea to realize our world  
as the effective field theory of light localized 
fields on the brane.
Supersymmetry (SUSY) is combined with this idea by considering 
the BPS solitons in SUSY field theories. 
They typically break the half of the original SUSY 
spontaneously. 
BPS domain walls are investigated in detail 
in $D=4$, ${\mathcal N}=1$ SUSY theories (with four supercharges),
on which effective field theories have 
two unbroken supercharges.
However in order to realize $D=4$, ${\mathcal N}=1$ SUSY 
theory on the world-volume, 
we need a higher-dimensional theory 
with eight supercharges.
In these theories scalar multiplets are 
hypermultiplets. 
They must parameterize curved hyper-K\"ahler 
(HK) manifolds~\cite{AF1,AF2} 
with a scalar potential admitting 
at least two discrete degenerate vacua.
These models are called the massive 
HK nonlinear sigma models (NLSM). 

In this talk, the HK quotient method~\cite{LR,HKLR} 
to construct HK manifolds is shown to be generalized 
to the massive models 
and the BPS domain wall in the simplest case 
of the Eguchi-Hanson target space is given. 
Keeping essential properties of eight supercharges,
we discuss a simpler and familiar case of 
$D=4$, ${\mathcal N}=2$ SUSY theories.
Also we use the ${\mathcal N}=1$ superspace formalism. 
A fully off-shell ${\mathcal N}=2$ superspace 
(the Harmonic superspace) formalism is discussed in 
the original paper~\cite{ANS}.

%%%%%%%%%%%%%%%%%%%%%%%%%%%%%%%%%%%%%%%%%%%%%

\section{K\"ahler Sigma Models and Walls}
Here we recall BPS walls in 
${\mathcal N}=1$ SUSY theories,
because we formulate ${\mathcal N}=2$ SUSY models
in terms of ${\mathcal N}=1$ superfields in the following sections.
Scalar fields belong to chiral superfields 
$\Phi^i(y,\theta) 
= \phi^i(y) + \sqrt {2} \theta \psi^i(y) + \theta\theta F^i(y)$
with $y^{\mu} = x^{\mu} + i \theta \sig^{\mu} \thb$.
The Lagrangian for the most general 
${\mathcal N}=1$ SUSY Lagrangian for chiral superfields 
(the generalized Wess-Zumino model) is given by~\cite{WB}
\beq
 {\mathcal L} 
 &=& \int d^4 \theta K(\Phi,\Phi\dagg) 
  + \left[ \int d^2\theta W(\Phi) + {\rm c.c.} \right] 
\eeq
with $K$ and $W$ real and holomorphic functions, 
called the K\"ahler potential and the superpotential, 
respectively. 
After elimination of auxiliary fields by their equations of motion 
$F^i = - g^{ij^*} \del_{j^*}W^* + \mbox{fermions}$, 
the bosonic part is calculated as
\beq
 {\mathcal L}_{\rm boson}  
 = - g_{ij^*} \del_{\mu}\phi^i \del^{\mu}\phi^{*j}  
   - g^{ij^*} \del_i W \del_{j^*} W^* \;
\eeq
with $g_{ij^*} \equiv 
\del_i \del_{j^*} K$ the K\"ahler metric. 
(We denote $\del_i = {\del \over \del \phi^i}$.)
The target manifold must be a K\"ahler manifold and 
therefore these models are called K\"ahler sigma models.

Assuming a domain wall configuration perpendicular to 
the third axis $x^3 = z$,
its energy density per unit area in the $x$-$y$ plane is given by
\beq
 E &=& \int d z  
 ( g_{ij^*} \del_z \phi^i \del_z \phi^{*j}
   + g^{ij^*} \del_i W \del_{j^*} W^* ) \non
 &=& \int d z  
       |\del_z \phi^i - e^{i\alpha} g^{ik^*} \del_{k^*} W^*|^2
   + \int d z ( e^{i\alpha} \del_z \phi^i \del_i W  + {\rm c.c.} )\non
 &\geq& \int d z ( \del_z \phi^i \del_i W   + {\rm c.c.} ) 
 = 2 {\rm Re} (e^{i\alpha} \Delta W) \;
\eeq
with the norm defined by $|V^i|^2 \equiv g_{ij^*}V^iV^{*j}$, 
$\Delta W \equiv W|_{z=\infty} - W|_{z=-\infty}$ 
and $\alpha$ arbitrary real constant.
Since we obtain the best bound at
$e^{-i\alpha} = \Delta W/|\Delta W|$, 
we derive the BPS bound 
$E \geq 2 |\Delta W|$ saturated by 
solutions of the BPS equation 
\beq
 \del_z \phi^i = e^{-i\alpha} g^{ij^*} \del_{j^*} W^*, \hs{5}
 e^{-i\alpha} = \Delta W/|\Delta W| \,.
  \label{BPSeq}
\eeq 
The SUSY transformation on the fermion $\psi^i$ 
with a BPS wall background is calculated as
$\delta_{\epsilon} \psi^i 
= \sqrt{2} (i \sig^{\mu} \bar{\epsilon} \del_{\mu} \phi^i 
+ \epsilon F^i ) 
= \sqrt {2} (i \sig^{z} \bar{\epsilon} e^{-i\alpha} - \epsilon) 
g^{ij^*}\del_{j^*}W^*$. 
Therefore two SUSYs satisfying 
$i e^{-i\alpha} \sig^{z} \bar{\epsilon}  = \epsilon$ 
out of four  are preserved and so 
the solutions are called $1/2$ BPS states.

The BPS domain walls (and their junction) in 
$D=4$, ${\mathcal N}=1$ SUSY NLSMs including runaway vacua
and singularity of the metric were discussed~\cite{NNS}.
It is easy to find ${\mathcal N}=1$ SUSY models 
admitting wall solutions, 
because $K$ and $W$ are 
arbitrary and independent to each other, 
but they are not for ${\mathcal N}=2$ SUSY 
as we will see in the following sections.

%%%%%%%%%%%%%%%%%%%%%%%%%%%%%%%%%%%%%%%%%%%%%

\section{Massive Hyper-K\"ahler Sigma Models and Walls}

We discuss hypermultiplets with potential terms. 
The on-shell component Lagrangian for massive HK model 
is well known~\cite{AF2}, whose bosonic part is 
\beq
 {\mathcal L}_{\rm boson} 
 = - g_{ij^*} \del_{\mu}\phi^i \del^{\mu}\phi^{*j} 
    - |\mu|^2 g_{ij^*}  k^i k^{*j} \;
\eeq
with $g_{ij^*}$ the target HK metric, 
$\mu$ a complex mass parameter and
$k^i(\phi,\phi^*)$ a tri-holomorphic Killing vector 
on the target HK manifold.\footnote{
The potential term can be interpreted by 
the Sherk-Schwarz (SS) dimensional reduction~\cite{SS} 
from six-dimensions, 
where HK sigma models are massless. 
Then the SS reduction to four (five) space-time dimensions 
is defined by 
$-i(\partial_5+i \partial_6)\phi^i=\mu k^i$ 
with $\mu \in {\bf C}$ 
($\partial_6 \phi^i=\mu k^i$ with $\mu \in {\bf R}$). 
}
Therefore an isometry on the manifold is required 
for a nontrivial potential to exist 
and vacua are given by fixed points of its action. 

We take $\mu$ real because phase can be absorbed into 
the definition of $k^i$.
Energy density for a wall 
perpendicular to the $z$-axis is
\beq
 E 
 &=& \int d z  
    ( g_{ij^*} \del_z \phi^i \del_z \phi^{*j}
      + \mu^2 g_{ij^*} k^i k^{*j} ) \non
 &=& \int d z |\del_z \phi^i - \mu k^i|^2
          %           (\del_z \phi^{*j} - \mu k^{*j})
   + \int d z ( \mu g_{ij^*} k^i \del_z \phi^{*j}   + {\rm c.c.} )\non
 &\geq& \int d z ( \mu g_{ij^*} k^i \del_z \phi^{*j}   + {\rm c.c.} ) 
 =  \mu \Delta D
\eeq
where $k^i = g^{ij^*} \del_{j^*} D$ with
$D(\phi,\phi^*)$ a real function called the Killing potential 
(moment map)\footnote{
It is in general difficult to find $D$ 
for given manifold and Killing vector.
} 
and $\Delta D \equiv D|_{z= \infty} - D|_{z= - \infty}$.
We thus obtain the BPS bound 
$E \geq \mu \Delta D$ saturated by 
the BPS equation $\del_z \phi^i = \mu k^i$~\cite{AT}. 

The Eguchi-Hanson space $T^*{\bf C}P^1$ 
admits $SU(2)$ tri-holomorphic isometry, 
one of whose three generators, say $\sig_3$, 
can be used to obtain the potential. 
A rotation around the third axis 
on the base $S^2 \simeq {\bf C}P^1$ has two fixed points 
on the North and South poles, both of which are vacua.
The BPS domain wall interpolating these vacua was firstly 
obtained by Abraham and Townsend~\cite{AT} 
in the component formalism. 
Lots of interesting BPS solitons were constructed 
in toric HK manifolds~\cite{GPTT,TL}.

%%%%%%%%%%%%%%%%%%%%%%%%%%%%%%%%%%%%%%%%%%%%%%%%%%%
\section{Massive Hyper-K\"ahler Models from 
${\mathcal N}=2$ SUSY QCD}

Let $(\Phi,\Psi)$ be ${\mathcal N}=2$ hypermultiplets 
with $\Phi$ and $\Psi$ being 
$N \times M$ and $M \times N$
matrix chiral superfields. 
To obtain nontrivial vacua we need an $U(M)$ gauge symmetry
introducing ${\mathcal N}=2$ vector multiplets 
$(V,\Sigma)$ with $V$ an $M \times M$  matrix vector superfield
and $\Sigma$ an $M \times M$ matrix chiral superfield. 
We work out for the $U(M)$ gauge group in which 
$U(1)$ part is essential to obtain discrete vacua.
We consider the Higgs branch of the theory 
taking the strong coupling limit $g\to \infty$ 
of gauge interactions, which eliminates 
the kinetic terms for $V$ and $\Sig$.
The gauge invariant Lagrangian is given by
\beq
&& {\mathcal L} = \int d^4 \theta
 \left[ \tr (\Phi\dagg\Phi e^V )  
 + \tr (\Psi\Psi\dagg e^{-V}) - c \, \tr V \right]  \non
&&\hs{5} + \left[ \int d^2\theta \,
  \big\{
  \tr \{ \Sigma (\Psi \Phi - b {\bf 1}_M) \} 
    +  {\sum_{a=1}^{N-1} m_a \tr (\Psi H_a \Phi)} \big\} 
         + {\rm c.c.}\right] ,
\label{linear}
\eeq
with $b \in {\bf C}$ and $c \in {\bf R}$ 
constituting a triplet of the Fayet-Iliopoulos parameters, 
$m_a$ complex mass
and $H_a$ Cartan generators of $SU(N)$.\footnote{
Flavor symmetry   
$\Phi \to \Phi' = g \Phi$,
$\Psi \to \Psi' = \Psi g^{-1}$ with 
$g \in SU(N)$ in the massless limit $m_a =0$
is explicitly broken by the mass to 
its Cartan $U(1)^{N-1}$ generated by $H_a$.
}
Eliminating superfields $V$ and $\Sig$ using 
their algebraic equations of motion, 
we obtain the Lagrangian in terms of independent superfields,
in which the K\"ahler potential is
\beq
 &&K = c\, \tr \sqrt{{\bf 1}_M + {4\over c^2} \Phi\dagg\Phi \Psi\Psi\dagg} 
   - c\, \tr \log \left( {\bf 1}_M 
    + \sqrt{{\bf 1}_M + {4\over c^2} \Phi\dagg\Phi \Psi\Psi\dagg}\right) \non
 && \hs{10}
    + c\, \tr \log \Phi\dagg\Phi  \;, \label{kahler}
\eeq
with a gauge fixing\footnote{
We discuss the $b \neq 0$ case here. 
The $b=0$ case must be discussed independently~\cite{ANS}.
}
\beq
 \Phi = \begin{pmatrix}
          {\bf 1}_M \cr \ph
        \end{pmatrix}  
        Q \;, \hs{5} 
 \Psi = Q ({\bf 1}_M, \psi) \;, \hs{5}
 Q = \sqrt b ({\bf 1}_M + \psi\ph)^{-\1{2}} \;, 
 \label{fixing2}
\eeq
with $\ph$ ($\psi$) an $(N-M) \times M$ [$M \times (N-M)$] matrix 
chiral superfield, 
and the superpotential is
\beq
 W = b \sum_a m_a \tr \left[
    H_a \begin{pmatrix}
          {\bf 1}_M \cr \ph
        \end{pmatrix} 
    ({\bf 1}_M + \psi\ph)^{-1}
    ({\bf 1}_M, \psi)  
   \right] \;. \label{superpot2}
\eeq
This is the massive extension of the HK NLSM on 
the cotangent bundle over the Grassmann manifold, 
$T^* G_{N,M}$, found by Lindstr\"om and Ro\v{c}ek~\cite{LR}.
This model contains 
${}_N C_M = N!/M! (N-M)!$ discrete degenerate vacua 
corresponding to independent gauge fixing 
conditions (\ref{fixing2})~\cite{ANS}.

%%%%%%%%%%%%%%%%%%%%%%%%%%%%%%%%%%%%%%%%%%%%%%%%%%%
\section{Domain Wall Solutions}
As seen in Section 2, the tension of the BPS domain wall 
is given by superpotential. 
This fact implies that we should take $b \neq 0$ 
in ${\mathcal N}=1$ superfields. 
The $M=1$ case of $U(1)$ gauge symmetry 
reduces to $T^* {\bf C}P^{N-1}$ with the superpotential, 
which admits $N$ parallel domain walls~\cite{TL}. 
Moreover if we take $N=2$ and $M=1$, the target space 
$T^*{\bf C}P^1$ is the Eguchi-Hanson space with the superpotential 
$W = b {\mu \over 1 + \ph \psi}$ ($\mu \equiv m_1$).
The BPS eq. (\ref{BPSeq}) in ${\mathcal N}=1$ superfields 
can be solved to give~\cite{ANNS}
\beq
 \ph = \psi^* = e^{|\mu| (z - z_0)} e^{i \delta}\;,    
 \label{solution} 
\eeq
where $z_0$ and $\delta$ are integral constants. 
They correspond to zero modes 
arising from spontaneously broken 
translational invariance perpendicular 
to domain wall configuration and 
$U(1)$ isometry $\sig_3$ in the {\it internal} space.

%%%%%%%%%%%%%%%%%%%%%%%%%%%%%%%%%%%%%%%%%%%%%%%%%%%%%%%%%%%%%
%                                                           %
% You may repeat \section{SECTION N-th HEADING TYPE HERE}   %
%                                                           %
% Do start a subsection or sub-subsection, do this:-        %
%                                                           %
%   \subsection{SUBSECTION HEADING TYPE HERE}               %
%                                                           %
%   \subsubsection{SUBSUBSECTION HEADING TYPE HERE}         %
%                                                           %
% instead of the above                                      %
%                                                           %
%%%%%%%%%%%%%%%%%%%%%%%%%%%%%%%%%%%%%%%%%%%%%%%%%%%%%%%%%%%%%

\section{Conclusions}
We have constructed massive HK NLSM 
on the cotangent bundle over $G_{N,M}$ in 
${\mathcal N}=1$ superfields,
which is the massive extension of Lindstr\"om and Ro\v{c}ek.
This model contains ${}_N C_M = N!/M! (N-M)!$ 
discrete degenerate vacua. 
A BPS Domain wall solution in the simplest $T^* {\bf C}P^1$ 
has been given. 

Constructing domain walls in non-Abelian gauge group 
remains as an interesting future work. 
Turning on the gauge coupling does not change vacua. 
BPS walls should also be similar as shown 
in the $M=1$ case~\cite{TL,KSIOS}.
Coupling to supergravity is possible as  
in the $M=1$ case~\cite{sugra}.

%%%%%%%%%%%%%%%%%%%%%%%%%%%%%%%%%%%%%%%%%%%%%%%%%%%%%%%%%%%%%
% Doing Acknowledgement                                     %
%%%%%%%%%%%%%%%%%%%%%%%%%%%%%%%%%%%%%%%%%%%%%%%%%%%%%%%%%%%%%

\section*{Acknowledgments}
We would like to thank
Masashi Naganuma for discussions in the early stage 
of this work. 
M.~N. is grateful to the organizers in 
QTS3 and the Argonne brane workshop.
This work is supported in part by Grant-in-Aid 
 for Scientific Research from the Japan Ministry 
 of Education, Science and Culture  13640269  (NS). 
The work of M.~N. was supported by the U.~S. Department
 of Energy under grant DE-FG02-91ER40681 (Task B).

\if0 %%%
%%%%%%%%%%%%%%%%%%%%%%%%%%%%%%%%%%%%%%%%%%%%%%%%%%%%%%%%%%%%%
% Doing Appendix(ices)                                      %
%%%%%%%%%%%%%%%%%%%%%%%%%%%%%%%%%%%%%%%%%%%%%%%%%%%%%%%%%%%%%

\appendix

\section{HEADING FOR APPENDIX A}

\renewcommand{\theequation}{A.\arabic{equation}}

TYPE TEXT FOR APPENDIX A HERE.

\section{HEADING FOR APPENDIX B}

\renewcommand{\theequation}{B.\arabic{equation}}

TYPE TEXT FOR APPENDIX B HERE.
\fi %%%

%%%%%%%%%%%%%%%%%%%%%%%%%%%%%%%%%%%%%%%%%%%%%%%%%%%%%%%%%%%%%
% Doing references:                                         %
%%%%%%%%%%%%%%%%%%%%%%%%%%%%%%%%%%%%%%%%%%%%%%%%%%%%%%%%%%%%%

\end{document}